\documentstyle[epsfig,preprint,aps,pra]{revtex}
\begin{document}

\tighten
\title{Near-Threshold  Photoproduction of $\eta$-mesons on
        Three-Nucleon Nuclei }
\author{
        N. V. Shevchenko$^{1}$, V. B. Belyaev$^{1,2}$,
    S. A. Rakityansky$^{2}$, S. A. Sofianos$^2$, W.~Sandhas$^3$}
\address{
         $^1$Joint Institute  for Nuclear Research, Dubna, 141980,
         Russia}
\address{
         $^2$Physics Department, University of South Africa,
         P.O. Box 392, Pretoria 0003, South Africa}
\address{
         $^3$Physikalisches Institut, Universit\H{a}t Bonn,
         D-53115 Bonn, Germany}
\maketitle
%%%%%%%%%%%%%%%%%%%%%%%%%%%%%%%%%%%%%%%%%%%%%%%%%%%%%%%%%%%%%%%%%%%%%%%%
\begin{abstract}
A microscopic few-body description of near-threshold coherent
photoproduction of the $\eta$ meson on tritium and $^3$He
targets is given. The photoproduction cross-section is
calculated using the Finite Rank Approximation (FRA) of the
nuclear Hamiltonian.  The results indicate a strong final state
interaction of the $\eta$ meson with the residual nucleus.
Sensitivity of the results to the choice of the $\eta N$
$T$-matrix is investigated. 

\end{abstract}
%%%%%%%%%%%%%%%%%%%%%%%%%%%%%%%%%%%%%%%%%%%%%%%%%%%%%%%%%%%%%%%%%%%%%%%%

\section{Introduction}
%%%%%%%%%%%%%%%%%%%%%%%

Investigations of the $\eta$-nucleus interaction are motivated
by various reasons. Some of them, such as the possibility of
forming quasi-bound states or resonances \cite{Heider} in the
$\eta$-nucleus system, are purely of nuclear nature. The others
are related to the study of the properties and structure of the
$S_{11}(1535)$ resonance which is strongly coupled to the $\eta
N$ channel.

For example, it is interesting to investigate the behavior of
the $\eta$-meson in nuclear media where, after colliding with
the nucleons, it readily forms the $S_{11}$ resonance.  The
interaction of this resonance with the surrounding nucleons can
be described in different ways \cite{Frank}, depending on
whether the structure of this resonance is defined in terms of
some quark configurations or by the coupling of meson-baryon
channels, as suggested in Ref. \cite{Keiser,Nieves}. The
estimation by Tiwari {\em et al.} \cite{Tiwari} shows, that
in case of pseudoscalar $\eta NN$ coupling there is an essential
density dependent reduction of the $\eta$-meson mass and of the
$\eta-\eta'$ mixing angle.

The importance of the influence of the nuclear medium on the
mesons passing through it, was recently emphasized by Drechsel
{\em et al.} \cite{Drech}. If this influence is described in
terms of self-energies and effective masses, then the passing
of $\pi$-mesons through the nucleus provides "saturation" of the
isobar propagator (or self-energy). This phenomenon
manifests itself even in light nuclei \cite{Drech}. Similar
ideas were discussed also in Ref. \cite{Fix1}. In other words,
the propagation of $\eta$-mesons inside the nucleus is a new
challenge for theorists.

Another interesting issue related to the $\eta$-nucleus
interaction is the study of charge symmetry breaking, which may
partly be attributed to the $\eta-\pi^0$ mixing (see, for
example, Refs. \cite{Coon,Wilkin,Mag,Ceci}).  In principle, one
can extract the value of the mixing angle from experiments
involving $\eta$-nucleus interaction and compare the results
with the predictions of quark models.  However, to do such an
extraction, one has to make an extrapolation of the
$\eta$-nucleus scattering amplitude into the area of unphysical
energies below the $\eta$-nucleus threshold. This is a highly
model dependent procedure requiring a reliable treatment of the
$\eta$-nucleus dynamics.

In this respect, few-body systems such as $\eta d$, $\eta$\,${}^3$He,
and $\eta$\,${}^4$He, have obvious advantages since they can be
treated using rigorous Faddeev-type equations. 
To the best of our knowledge, the exact AGS theory \cite{AGS}
has been used in the few calculations (see Refs. 
\cite{Ueda,Ours1,Fix2,Fix3}) for the $\eta d$ and in one
recent calculation \cite{Fix_He} for the $\eta {}^3$H and 
$\eta {}^3$He systems.

A solution of the few-body equations presupposes the knowledge of the
corresponding two-body $T$-matrices $t_{\eta N}$ and $t_{NN}$ off the
energy shell.  Due to the fact that at low energies the $\eta$ meson
interacts with a nucleon mainly via the formation of the
$S_{11}$-resonance, the inclusion of the higher partial waves
($\ell>0$) is unnecessary.  Furthermore, since the $\eta N$
interaction is poorly known, the effect of the fine tuned details of
the ``realistic'' $NN$ potentials would be far beyond the level of the
overall accuracy of the $\eta A$ theory.  

In contrast to the well-established $NN$ forces, the $\eta N$
interaction is constructed using very limited information available,
namely, the $\eta N$ scattering length and the parameters of the
$S_{11}$-resonance.  Furthermore, only the resonance parameters are
known more or less accurately while the scattering length (which is
complex) is determined with large uncertainties. Moreover, practically
nothing is known about the off-shell behavior of the $\eta N$
amplitude.
It is simply assumed that the off-shell behavior of this
amplitude could be approximated (like in the case of $\pi$ mesons) by
appropriate Yamaguchi form-factors (see, for example,
Refs. \cite{Ueda,Ours1,Fix2,Fix3,Gar,Deloff}).
However, if the available data are used to construct a potential via,
for example, Fiedeldey's inverse scattering procedure \cite{Fied}, the
resulting form factor of the separable potential is not that simple.
The problem becomes even more complicated due to the multichannel
character of the $\eta N$ interaction with the additional off-shell
uncertainties stemming from the $\pi$-meson channel.

In such a situation, it is desirable to narrow as much as possible the
uncertainty intervals for the parameters of $\eta N$ interaction.
This could be done by demanding consistency of theoretical
predictions based on these parameters, with  existing experimental
data for two-, three-, and four-body $\eta$-nucleus processes.  This
is one of the objectives of the present work. To do this, we calculate
the cross sections of coherent $\eta$-photoproduction on $^3$He and
$^3$H and study their sensitivity to the parameters of the $\eta N$
amplitude.

%%%%%%%%%%%%%%%%%%%%%%%%%%%%%%%%%%%%%%%%%%%%%%%%%%%%%%%%%%%%%%%%%%%%%%
\section{Formalism}
%%%%%%%%%%%%%%%%%%%
We start by assuming that the Compton scattering on a nucleon,
$$
       \gamma + N \rightarrow N + \gamma\ ,
$$
as well as the processes of multiple re-appearing of the photon in the
intermediate states,
$$
   \gamma + N\rightarrow N+\eta\rightarrow \gamma +N\rightarrow
   N+\eta\rightarrow\dots\ ,
$$
give a negligible contribution to the coherent
$\eta$-photoproduction on a nucleus $A$. 
In our model, we also neglect virtual excitations and breakup of the nucleus
immediately after its interaction with the photon. With these
assumptions, the process
\begin{equation}
\label{gAAeta}
        \gamma + A \rightarrow A + \eta
\end{equation}
can be formally described in two steps: in the first one, the
photon produces the $\eta$ meson on one of the nucleons,
\begin{equation}
\label{gNNeta}
        \gamma + N \rightarrow N + \eta\ ,
\end{equation}
in the second step (final state interaction) the $\eta$ meson is
elastically scattered off the nucleus,
\begin{equation}
\label{etaAAeta}
        \eta + A \rightarrow A + \eta\ .
\end{equation}
An adequate treatment of the scattering step is, of course, the
most difficult and crucial part of the theory. 
The first microscopic calculations concerning the low-energy
scattering of the $\eta$-meson from ${}^3$H, ${}^3$He, and ${}^4$He
were done in
Refs.~\cite{Ours2,Ours21,Ours22,Ours23,Ours24,Ours25,Ours26} where the
few-body dynamics of these systems was treated by employing the
Finite-Rank Approximation (FRA) \cite{FRA} of the nuclear
Hamiltonian. This approximation consists in neglecting the continuous
spectrum in the spectral expansion $$ H_A=\sum_n{\cal
E}_n|\psi_n\rangle\langle\psi_n|+ \mbox{continuum}
$$
of the Hamiltonian $H_A$ describing the nucleus. Since the three- and
four-body nuclei have only one bound state, FRA reduces to
\begin{equation}
\label{FRAapprox}
   H_A\approx{\cal E}_0|\psi_0\rangle\langle\psi_0|\ .
\end{equation}
Physically, this means that we exclude the virtual excitations
of the nucleus during its interaction with the $\eta$ meson. It
is clear that the stronger the nucleus is bound, the smaller is
the contribution from such processes to the elastic $\eta A$
scattering.  By comparing with the results of the exact AGS
calculations, it was shown\cite{AGSetad} that even for $\eta d$
scattering, having the weakest nuclear binding, the FRA method
works reasonably well, which implies that we obtain sufficiently
accurate results by applying this method to the $\eta$\,${}^3$H,
$\eta$\,${}^3$He, and even more so to the $\eta$\,${}^4$He
scattering.

In essence, the FRA method can be described as follows (for details see
Ref.\cite{FRA}). Let
$$
    H=h_0+V+H_A
$$
be the total $\eta A$ Hamiltonian, where $h_0$ describes the free
$\eta$-nucleus motion and
$$
     V=\sum_{i=1}^AV_i
$$
the sum of the two-body $\eta$-nucleon potentials. The
Lippmann-Schwinger equation
\begin{equation}
\label{LSinitial}
   T(z)=\sum_{i=1}^AV_i+\sum_{i=1}^AV_i(z-h_0-H_A)^{-1}T(z)
\end{equation}
for the $\eta$-nucleus $T$-matrix can be rewritten as
\begin{equation}
\label{Teq}
   T(z)=W(z)+W(z)M(z)T(z)\ ,
\end{equation}
where
\begin{eqnarray}
\label{kernel}
   M(z) &=& G_0(z)H_AG_A(z)\ ,\\
\label{G0}
   G_0(z) &=& (z-h_0)^{-1}\ ,\\
\label{Ga}
   G_A(z) &=& (z-h_0-H_A)^{-1}\ ,
\end{eqnarray}
and the auxiliary operator $W(z)$ is split into $A$ components of
Faddeev-type,
\begin{equation}
\label{T0sum}
   W(z) = \sum_{i=1}^AW_i(z)\ ,
\end{equation}
satisfying the following system of equations
\begin{equation}
\label{T0i}
   W_i(z) = t_i(z)+t_i(z)G_0(z)\sum_{j\ne i}^AW_j(z)\\
\end{equation}
with $t_i$ being the two-body $T$-matrix describing the interaction
of the
$\eta$-meson with the $i$-th nucleon, {\it i.e.},
\begin{equation}
\label{tetaNi}
   t_i(z) = V_i+V_iG_0(z)t_i(z)\ .
\end{equation}
It should be emphasized that up to this point no approximation
has been made and, therefore, the set of equations
(\ref{Teq}-\ref{tetaNi}) is equivalent to the initial
equation~(\ref{LSinitial}). However, to solve Eq. (\ref{Teq}),
we have to resort to the approximation (\ref{FRAapprox}) which
simplifies its kernel (\ref{kernel}) to
\begin{equation}
\label{kernelapprox}
    M(z)\approx
    \frac{{\cal E}_0|\psi_0\rangle\langle\psi_0|}
    {(z-h_0)(z-{\cal E}_0-h_0)}\ .
\end{equation}
With this approximation, the sandwiching of Eq. (\ref{Teq})
between $\langle\psi_0|$ and $|\psi_0\rangle$ and the partial wave
decomposition give a one-dimensional integral equation for the
amplitude of the process (\ref{etaAAeta}). 
Although this one-dimensional equation may look similar to the
integral equation of the first-order optical-potential theory, the FRA
approach is quite different. Firstly, in contrast to the optical
potential of the first order, the operator $W(z)$ includes all orders
of rescattering via solution of Eq. (\ref{T0i}). Secondly, the 
$\eta N$ amplitudes $t_i(z)$ entering Eq. (\ref{T0i}), are taken as
operators in the many-body space and off the energy shell (note that
$G_0$ in Eq. (\ref{tetaNi}) is four-body propagator with $\eta A$
reduced mass and $z$ is total four-body energy), i.e.  the FRA method
does not involve the so-called ``impulse approximation'' (using free
two-body amplitudes for $t_i$) which is an indispensable part of the
optical theory.

The question then arises how a photon can be included in
this formalism in order to describe the photoproduction process
(\ref{gAAeta}). This can be achieved by following the same
procedure as in Ref.\cite{Ours3} where the reaction
(\ref{gAAeta}) with $A=2$ was treated within the framework of
the exact AGS equations, and the photon was introduced by
considering the $\eta N$ and $\gamma N$ states as two different
channels of the same system. This implies that the operators
$t_i$ should be replaced by $2 \times 2$ matrices. It is clear
that such replacements of the kernels of the integral
equations~(\ref{T0i}) and subsequently of the integral equation
(\ref{Teq}) lead to solutions having a similar matrix form
\begin{equation}
\label{matrices}
t_{i} \to \left(
\begin{array}{cc}
 t_i^{\gamma \gamma} & t_i^{\gamma \eta} \\
 t_i^{\eta \gamma}   & t_i^{\eta \eta}
\end{array}
\right) \Longrightarrow\quad
W_i \to \left(
\begin{array}{cc}
 W_{i}^{\gamma \gamma} & W_{i}^{\gamma \eta} \\
 W_{i}^{\eta \gamma}   & W_{i}^{\eta \eta}
\end{array}
\right)\Longrightarrow\quad
T \to \left(
\begin{array}{cc}
 T^{\gamma \gamma} & T^{\gamma \eta} \\
 T^{\eta \gamma}   & T^{\eta \eta}
\end{array}
\right)\ .
\end{equation}
Here $t_i^{\gamma \gamma}$ describes the Compton scattering,
$t_i^{\eta \gamma}$ the photoproduction process, and $t_i^{\eta \eta}$
the elastic $\eta$ scattering on the $i$-th nucleon.  What is finally
needed is the cross section
\begin{equation}
\label{sechenie}
      \frac{{\rm d}\sigma}{{\rm d}\Omega} =
      \frac{\mu_{\eta A}}{(2 \pi)^2} \, \frac{k_{\eta}}{k_{\gamma}} \,
      \frac{E_\gamma m_A}{E_\gamma+m_A} \,
      \frac14\sum_{s'_z,s_z,\epsilon}
      \left|
      \langle\vec{k}_\eta,\varphi\phi^a_{s'_z,t_z}|
      T^{\eta\gamma}({\cal E}_0+E_\gamma)
      |\varphi\phi^a_{s_z,t_z},\vec{k}_\gamma,\epsilon\rangle
      \right|^2
\end{equation}
of the reaction (\ref{gAAeta}) 
averaged over orientations $s_z$ of the
initial nuclear spin and photon polarization $\epsilon$ and
summed over spin orientations $s'_z$ in the final state.  Here 
$\varphi$ and $\phi^a_{s_z,t_z}$ are the spatial and spin-isospin parts
of $\psi_0$ (with the third components of the nuclear spin and isospin
being $s_z$ and $t_z$ respectively),
$\vec{k}_\gamma$ and $\vec{k}_\eta$ are the momenta of the photon and
$\eta$ meson, $E_\gamma$ is the energy of the photon, $m_A$ the mass
of the nucleus, and $\mu_{\eta A}$ the reduced mass of the meson and
the nucleus.

However, it is technically more convenient to consider radiative
$\eta$-absorption, {\it i.e.}, the inverse reaction. Then the
photoproduction cross section can be obtained by applying the
principle of detailed balance. The reason for this is that all the
processes in which the photon appears more than once, {\it i.e.}, the
terms of the integral equations of type
$W^{\gamma\gamma}MT^{\gamma\eta}$ or $W^{\eta\gamma}MT^{\gamma\eta}$
involving more than one electromagnetic vertex, can be neglected.
Omission of these terms in (\ref{Teq}) results in decoupling the
elastic scattering equation
\begin{equation}
\label{elastic}
      T^{\eta\eta}=W^{\eta\eta}+W^{\eta\eta}MT^{\eta\eta}
\end{equation}
from the equation for the radiative absorption
\begin{equation}
\label{absorption}
      T^{\gamma\eta}=W^{\gamma\eta}+W^{\gamma\eta}MT^{\eta\eta}\ .
\end{equation}
Once $T^{\eta\eta}$ is calculated, the radiative absorption $T$-matrix
(\ref{absorption}) can be obtained by integration.

Therefore, the procedure of calculating the photoproduction
cross section (\ref{sechenie}) consists of the following steps:
\begin{itemize}
\item
Solving the system of equations
\begin{equation}
\label{proc1}
      W_i^{\eta\eta}=t_i^{\eta\eta}+t_i^{\eta\eta}G_0
      \sum_{j\ne i}^AW_j^{\eta\eta}
\end{equation}
for the auxiliary elastic-scattering operators $W_i^{\eta\eta}$\ .
\item
Calculating (by integration) the auxiliary matrices
$W_i^{\gamma\eta}$ from
\begin{equation}
\label{proc2}
      W_i^{\gamma\eta}=t_i^{\gamma\eta}+t_i^{\gamma\eta}G_0
      \sum_{j\ne i}^AW_j^{\eta\eta}\ .
\end{equation}
\item
Solving the integral equation
\begin{equation}
\label{proc3}
      T^{\eta\eta}=\sum_{i=1}^AW_i^{\eta\eta}+\sum_{i=1}^AW_i^{\eta\eta}
      MT^{\eta\eta}
\end{equation}
for the elastic-scattering $T$-matrix.
\item
Calculating (by integration) the radiative absorption $T$-matrix
\begin{equation}
\label{proc4}
      T^{\gamma\eta}=\sum_{i=1}^AW_i^{\gamma\eta}+
      \sum_{i=1}^AW_i^{\gamma\eta}MT^{\eta\eta}\ .
\end{equation}
\item
Substituting this $T$-matrix into Eq. (\ref{sechenie}) to obtain the
differential cross section for the photoproduction.  This is possible
because the absolute values of the photoproduction and radiative absorption
$T$-matrices coincide.
\end{itemize}

%%%%%%%%%%%%%%%%%%%%%%%%%%%%%%%%%%%%%%%%%%%%%%%%%%%%%%%%%%%%%%%%%%%%%%
\section{Two-body interactions}
%%%%%%%%%%%%%%%%%%%%%%%%%%%%%%%
To implement the steps described in the previous section,
we need the two-body $T$-matrices $t^{\eta\eta}$ and $t^{\gamma\eta}$
for the elastic $\eta N$ scattering and the radiative absorption
$N(\eta,\gamma)N$ on a single nucleon, respectively. Furthermore, all
equations (\ref{proc1}-\ref{proc4}) have to be sandwiched between
$\langle\psi_0|$ and $|\psi_0\rangle$ (ground state of
the nucleus).  Since at low energies both the elastic scattering and
photoproduction of the $\eta$ meson on a nucleon proceed mainly via formation
of the $S_{11}$ resonance, we may retain only the $S$-waves in the
partial wave expansions of the corresponding two-body $T$-matrices.

%%%%%%%%%%%%%%%%%%%%%%%%%%%%%%%%%%%%%%%%%%%%%%%%%%%%
\subsection{Elastic $\eta N$ scattering}
\label{twobodyint}
%%%%%%%%%%%%%%%%%%%%%%%%%%%%%%%%%%%%%%%%%%%%%%%%%%%%
The problem of constructing an $\eta N$ potential or directly the
corresponding $T$-matrix $t^{\eta\eta}$ has no unique solution since
the only experimental information available consists of the
$S_{11}$-resonance pole position $E_0-i\Gamma/2$ and the $\eta N$
scattering length $a_{\eta N}$.  In the present work we use three
different versions of $t^{\eta\eta}$.

%%%%%%%%%%%%%%%%%%%%%%%%%%%%%%%%%%%%%%%%%%%%%%%%%%%%%
\subsubsection{Version I}
%%%%%%%%%%%%%%%%%%%%%%%%%%%%%%%%%%%%%%%%%%%%%%%%%%%%%
Without any scattering data it is practically impossible to construct
a reliable $\eta N$ potential. 
In the low-energy region, however,  the elastic scattering can be viewed 
as the process of formation and subsequent decay of $S_{11}$ 
resonance, {\it i.e.},
\begin{equation}
\label{diagram}
        \eta + N\,\longrightarrow\,S_{11}\,
        \longrightarrow\,N+\eta\ .
\end{equation}
This implies that the corresponding Breit-Wigner formula could be a
good approximation for the $\eta N$ cross section.  Therefore, we may
adopt the following ansatz
\begin{equation}
    t^{\eta\eta}(k',k;z) = g(k') \, \tau(z) \, g(k)
\label{tetan}
\end{equation}
where the propagator $\tau(z)$, describing the intermediate state of
the process (\ref{diagram}), is assumed to have a simple Breit-Wigner
form
\begin{equation}
    \tau(z) = \frac {\lambda}{z - E_0 + i\Gamma/2}\ ,
\label{tau1}
\end{equation}
which guaranties that the $T$-matrix (\ref{tetan}) has a pole at the
proper place.  The vertex function $g(k)$ for the processes $\eta
N$\,$\leftrightarrow S_{11}$ is chosen to be
\begin{equation}
\label{ffactor}
               g(k)=(k^2+\alpha^2)^{-1}
\end{equation}
which in configuration space is of Yukawa-type. The range parameter
$\alpha = 3.316$\,fm$^{-1}$ was determined in Ref.~\cite{alpha} while
the parameters of the $S_{11}$-resonance
$$
        E_0=1535\,{\rm MeV}-(m_N+m_\eta)\ ,
        \qquad \Gamma=150\,{\rm MeV}
$$
are taken from Ref.~\cite{PDG}.  The strength parameter
$\lambda$ is chosen to reproduce the $\eta$-nucleon scattering
length $a_{\eta N}$,
\begin{equation}
     \lambda= 2 \pi \, \frac{\alpha^4(E_0-i\Gamma/2)}
        {\mu_{\eta N}}a_{\eta N}\,,
\label{t000}
\end{equation}
the imaginary part of which accounts for the flux losses into the $\pi
N$ channel. Here $\mu_{\eta N}$ is the $\eta N$ reduced mass.

The two-body scattering length $a_{\eta N}$ is not accurately
known. Different analyses~\cite{Batinic} provided values for $a_{\eta N}$
in the range
\begin{equation}
           0.27\ {\rm fm}\le{\rm Re\,}a_{\eta N}\le 0.98\
{\rm fm}\ ,\qquad
       0.19\ {\rm fm}\le{\rm Im\,}a_{\eta N}\le 0.37\ {\rm fm}\ .
\label{interval}
\end{equation}
In most recent publications the value used for Im $a_{\eta N}$
is around $0.3$\,fm.  However, for Re\,$a_{\eta N}$ the
estimates are still very different (compare, for example,
Refs.~\cite{Green97} and~\cite{Speth}).  In the present work we
assume that
\begin{equation}
\label{etaNlength}
        a_{\eta N} = (0.75 + i 0.27)\,{\rm fm}\ .
\end{equation}
The $T$-matrix $t^{\eta\eta}$ constructed in this way reproduces the
scattering length (\ref{etaNlength}) and the $S_{11}$ pole, but apparently
violates the two-body unitarity since it does not obey the two-body
Lippmann-Schwinger equation.

%%%%%%%%%%%%%%%%%%%%%%%%%%%%%%%%%%%%%%%%%%%%%%%%%
\subsubsection{Version II}
%%%%%%%%%%%%%%%%%%%%%%%%%%%%%%%%%%%%%%%%%%%%%%%%%
An alternative way of constructing the two-body $T$-matrix
$t^{\eta\eta}$ is to solve the corresponding Lippmann-Schwinger
equation with an appropriate separable potential having the same
form-factors (\ref{ffactor}).  However, a one-term separable
$T$-matrix obtained in this way, does not have a pole at
$z=E_0-i\Gamma/2$.  To recover the resonance behavior in this case,
we use the trick suggested in Ref.~\cite{Deloff}, namely, we use an
energy-dependent strength of the potential
$$
       V(k,k';z)=g(k)\,
         \left[ \Lambda + C \frac{\zeta}{\zeta-z} \right] \, g(k')
$$
where $\Lambda$ is complex while $C$ and $\zeta$ are real constants.
With this ansatz for the potential, the Lippmann-Schwinger equation
gives the $T$-matrix in the form~(\ref{tetan}) with
\begin{equation}
    \tau(z) = -\left(\frac{4 \pi \alpha^3}{\mu_{\eta N}}\right)
    \frac{\Lambda (\zeta-z) + C \zeta}
    {\zeta-z-\left[\Lambda (\zeta-z)
    + C \zeta\right] / (1 - i\sqrt{2 z \mu_{\eta N}} /\alpha)^2}\ .
\label{tdeloff}
\end{equation}
The constants $\Lambda$, $C$, and $\zeta$ can be chosen in such
a way that the corresponding scattering amplitude reproduces the
scattering length $a_{\eta N}$ and has a pole at
$z=E_0-i\Gamma/2$.
This version of $t^{\eta\eta}$ also reproduces the scattering length
(\ref{etaNlength}) and the $S_{11}$ pole.  Moreover, it is consistent
with the condition of two-body unitarity.

%%%%%%%%%%%%%%%%%%%%%%%%%%%%%%%%%%%%%%%%%%%%%%%%%
\subsubsection{Version III}
%%%%%%%%%%%%%%%%%%%%%%%%%%%%%%%%%%%%%%%%%%%%%%%%%
We can also construct $t^{\eta\eta}$ in the form (\ref{tetan}), with
the same $\tau(z)$ as in (\ref{tau1}), but obeying the unitarity condition
\begin{equation}
\label{unitarity}
        (1-2\pi it^{\eta\eta})(1-2\pi it^{\eta\eta})^\dag=1\ .
\end{equation}
Of course, with the simple form (\ref{tetan}) we cannot satisfy the
condition (\ref{unitarity}) at all energies. To simplify the
derivations, we impose this condition on $t^{\eta\eta}$ at $z=E_0$. 
%as it was suggested in~\cite{BohrMot}.  
Since Eq. (\ref{unitarity}) is real, it can fix only one parameter and 
we need one more condition to fix both the real and imaginary parts of 
the complex $\lambda$.  As the second equation, we used the 
real or imaginary part of Eq.~(\ref{t000}) 
(version III-a or III-b respectively) 
with $a_{\eta N}$ given by (\ref{etaNlength}).

This procedure guaranties two-body unitarity and gives the correct
position of the resonance pole, but the resulting $t^{\eta\eta}$
provides a value of $a_{\eta N}$ which, of course, slightly differs
from~(\ref{etaNlength}), namely,
\begin{eqnarray}
\label{etaNlength1a}
     a_{\eta N}=(0.77 +i0.26)\, \mbox{fm\ , \qquad version III-a}\ , \\
\label{etaNlength1b}
     a_{\eta N}=(0.79 +i0.32)\, \mbox{fm\ , \qquad version III-b}\ .
\end{eqnarray}

In what follows we use these three versions of the matrix
$t^{\eta\eta}$. All of them have the same separable form
(\ref{tetan}) but different $\tau(z)$.  Comparison of the results
obtained with these three $T$-matrices can give an indication of the
importance of two-body unitarity in the photoproduction
processes.

%%%%%%%%%%%%%%%%%%%%%%%%%%%%%%%%%%%%%%%%%%%%%%%%%%%%
\subsection{Radiative absorption $N(\eta,\gamma)N$}
%%%%%%%%%%%%%%%%%%%%%%%%%%%%%%%%%%%%%%%%%%%%%%%%%%%%
In constructing the radiative absorption $T$-matrix $t^{\gamma\eta}$, the
$S_{11}$ dominance in the near-threshold region also plays an
important role. It was experimentally shown~\cite{Krusche} that, at
low energies, the reaction (\ref{gNNeta}) proceeds mainly via
formation of the $S_{11}$-resonance, 
\begin{equation}
\label{diagramgamma}
        \gamma + N\,\longrightarrow\,S_{11}\,
        \longrightarrow\,N+\eta\ ,
\end{equation}
by the $S$-wave $E_{0^+}$ multipole.  This means that in the standard
CGLN decomposition of the $N(\gamma,\eta)N$ amplitude (see, for
example, Ref.\cite{alpha}) only the term proportional to the dot-product
$(\vec\sigma\cdot\vec\epsilon)$ of the nucleon spin and photon polarization
can be retained, i.e.
\begin{equation}
\label{CGLN}
     t^{\gamma \eta}=f^{\gamma \eta}(\vec\sigma\cdot\vec\epsilon)\ .  
\end{equation}

The dominance of the process (\ref{diagramgamma}) 
implies that $f^{\gamma \eta}$ in this energy region can be written in
a separable form similar to (\ref{tetan}). To construct such a
separable $T$-matrix, we use the results of Ref.~\cite{Green99} where
$t^{\gamma \eta}$ was considered as an element of a multi-channel
$T$-matrix which simultaneously describes experimental data for the
processes
\begin{eqnarray}
\nonumber
    &{}& \pi + N \to \pi + N,    \quad   \pi + N \to \eta + N,
\\
\nonumber
    &{}& \gamma + N \to \pi + N, \quad  \gamma + N \to \eta + N
\end{eqnarray}
on the energy shell in the $S_{11}$-channel 
(the $\eta N$ scattering length obtained in Ref.~\cite{Green99} is the 
same as we use for constructing versions I and II of $t^{\eta \eta}$).
In the present work, we take the $T$-matrix $t_{\rm on}^{\gamma
\eta}(E)$ from Ref.~\cite{Green99} and extend it off the energy shell
via
\begin{equation}
\label{toff}
    f_{\rm off}^{\gamma \eta}(k',k;E) =
    \frac{\kappa^2 + E^2}{\kappa^2 + {k'}^2} \,
        t_{\rm on}^{\gamma \eta}(E) \,
    \frac{\alpha^2 + 2 \mu_{\eta N} E}{\alpha^2 + k^2}\ ,
\end{equation}
where $\kappa$ is a parameter. The Yamaguchi form-factors used in this
ansatz go to unity on the energy shell.  Since $\kappa$ is not known,
this parameter is varied in our calculations within an
interval $1\,{\rm fm}^{-1}<\kappa<10\,{\rm fm}^{-1}$ which is a typical
range for meson-nucleon forces. It is known that $t^{\gamma \eta}$ is
different for neutron and proton.  In this work we assume that they
have the same functional form (\ref{toff}) but differ by a constant
factor,
$$
       t_{\rm n}^{\gamma \eta} = A \, t_{\rm p}^{\gamma\eta}\ .
$$
Multipole analysis~\cite{Muk} gives for this factor the
estimate $ A= -0.84 \pm 0.15$\,.
Therefore, if we direct the $z$-axis along the photon momentum 
$\vec k_\gamma$, the radiative absorption $T$-matrix entering our 
equations, can be written as
\begin{equation}
\label{general}
      t^{\gamma \eta}=f_{\rm off}^{\gamma \eta}\cdot
      \left(\sigma_x\epsilon_x+\sigma_y\epsilon_y\right)
      \left(P_p+AP_n\right)\ ,
\end{equation}
where $\epsilon_x$ and $\epsilon_y$ are transverse components of the
photon polarization vector while $P_p$ and $P_n$ are the operators
projecting onto the proton and neutron isotopic states respectively.

%%%%%%%%%%%%%%%%%%%%%%%%%%%%%%%%%%%%%%%%%%%%%%%%%%%%
\subsection{Nuclear subsystem}
%%%%%%%%%%%%%%%%%%%%%%%%%%%%%%%%%%%%%%%%%%%%%%%%%%%%
Since the $T$-matrices $t^{\eta\eta}$ and $t^{\gamma\eta}$ are poorly
known and their uncertainties significantly limit the overall accuracy
of the theory, it is not necessary to use any sophisticated
(``realistic'') potential to describe the $NN$ interaction. Therefore
we may safely assume that the nucleons interact with each other only
in the $S$-wave state.

To obtain the necessary nuclear wave function $\psi_0$, we solve the
few-body equations of the Integro-Differential Equation Approach
(IDEA)~\cite{idea1,idea2} with the Malfliet-Tjon potential~\cite{mt}.
This approach is based on the Hyperspherical Harmonic expansion method
applied to Faddeev-type equations. In fact, in the case of $S$-wave
potentials, the IDEA is fully equivalent to the exact Faddeev
equations.  Therefore, the bound states used in our calculations are
derived, to all practical purposes, via an exact formalism.

%%%%%%%%%%%%%%%%%%%%%%%%%%%%%%%%%%%%%%%%%%%%%%%%%%%%%%%%%%%%%%%%%%%%%
\section{Spin-isospin average}
\label{spinisospina}
%%%%%%%%%%%%%%%%%%%%%%%%%%%%%%%%%%%%%%%%%%%%%%%%%%%%%%%%%%%%%%%%%%%%%
The wave function $\psi_0=\varphi\phi^a_{s_z,t_z}$ of the $^3$H/$^3$He
system obtained by solving the IDEA equations with the Malfliet-Tjon
potential, has only the symmetric $S$-wave spatial component $\varphi$
multiplied by the antisymmetric spin-isospin part
$$
       \phi^a_{s_z,t_z}=\frac{1}{\sqrt{2}}\left(
       \chi'_{s_z}\eta''_{t_z}-\chi''_{s_z}\eta'_{t_z}\right)\ ,
$$
where $\chi'$, $\chi''$ and $\eta'$, $\eta''$ are the mixed symmetry
states in the spin and isospin sub-spaces. The matrix element
of $T^{\eta\gamma}$ in Eq. (\ref{sechenie}) involves the average not
only over the spatial part of $\psi_0$ but over $\phi^a$ as well.
The average $\langle\phi^a |T^{\eta\gamma|}|\phi^a\rangle$ can be done 
before we start solving equations (\ref{proc1}-\ref{proc4}).

Since $t^{\eta\eta}$, $G_0$, and $M$ do not involve spin-isospin operators,
the averaging of Eqs. (\ref{proc1}) and (\ref{proc3}) over $\phi^a$ is
trivial: It does not produce any additional coefficients.
Eq. (\ref{proc2}), however, changes. Indeed, for each nucleon
($i=1,2,3$), it involves the operator (\ref{general}) which causes
nucleon spin to flip over. 

Formal averaging of $t_j^{\gamma\eta}$ (for $j=1,2,3$) over the states
$\phi^a_{s_z,t_z}$ having definite values of the $z$-components  of 
total spin ($s_z$) and isospin ($t_z$), gives the same results
\begin{equation}
\label{avrg}
  \langle\phi^a_{s'_z,t_z}|t_j^{\gamma\eta}|\phi^a_{s_z,t_z}\rangle=
  \left\{
  \begin{array}{lcllc}
  \displaystyle
  \delta_{-s'_z,s_z}f_{\rm off}^{\gamma \eta}
  \cdot\frac{A}{3}(\epsilon_x+i\epsilon_y) &,&
  \quad \mbox{for } & t_z=+1/2\ & (^3{\rm He})\\[5mm]
  \displaystyle
  \delta_{-s'_z,s_z}f_{\rm off}^{\gamma \eta}
  \cdot\frac13(\epsilon_x+i\epsilon_y) &,&
  \quad \mbox{for } & t_z=-1/2\ & (^3{\rm H})
  \end{array}
  \right.
\end{equation}
for all three nucleons. This means that all three matrix elements
$\langle\phi^a_{s'_z,t_z}|W_j^{\gamma\eta}|\phi^a_{s_z,t_z}\rangle$
($j=1,2,3$) acquire the same coefficient, namely,
$A(\epsilon_x+i\epsilon_y)/3$ or
$(\epsilon_x+i\epsilon_y)/3$ depending on $t_z$.  Via
Eq. (\ref{proc4}), the same coefficient goes to the matrix element
$\langle\phi^a_{s'_z,t_z}|T^{\gamma\eta}|\phi^a_{s_z,t_z}\rangle$.
Since $\epsilon_x^2+\epsilon_y^2=1$, this gives the factor $|A|^2/9$
(for the case of $^3$He) or $1/9$ (for the case of $^3$H) in the final
Eq. (\ref{sechenie}) for the photoproduction cross section.
Thus, the cross section for the $^3$He target quadratically depends on
$A$ while for the case of $^3$H it is independent of $A$.

%%%%%%%%%%%%%%%%%%%%%%%%%%%%%%%%%%%%%%%%%%%%%%%%%%%%%%%%%%%%%%%%%%%%%
\section{Results and discussion}
\begin{figure}
\begin{center}
\unitlength=1mm
\begin{picture}(130,100)
\put(0,0){\epsfig{file=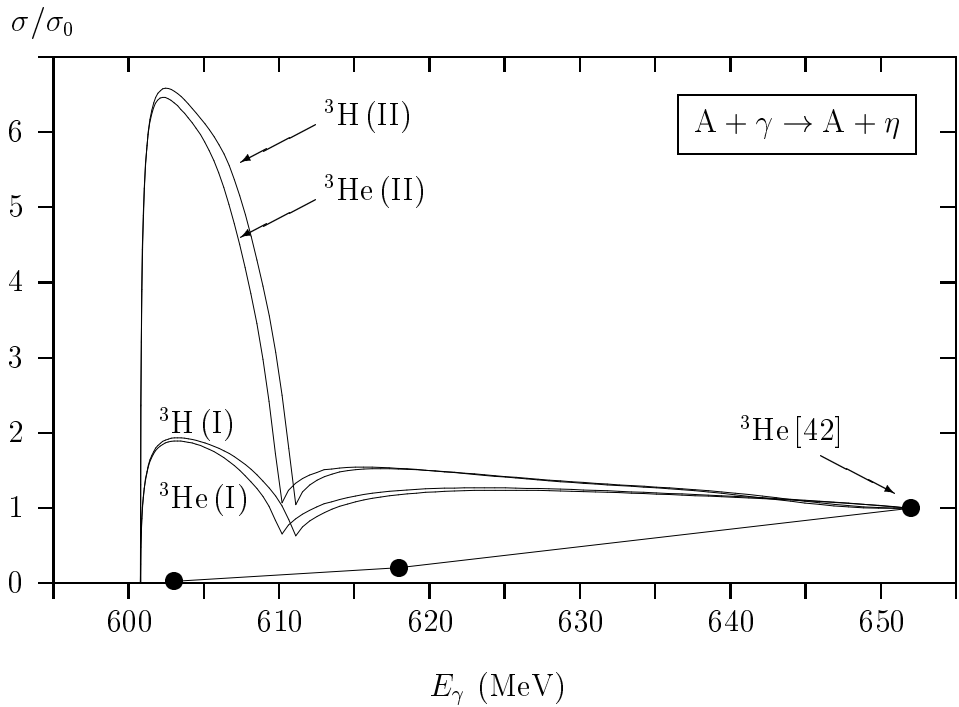}}
\end{picture}
\medskip
\caption{ 
Normalized cross section of the coherent
$\eta$-photoproduction on the $^3$H and $^3$He targets, calculated
with the two versions of $t^{\eta\eta}$ which are denoted as (I) and
(II) respectively. All curves correspond to $a_{\eta
N}=(0.75+i0.27)\,{\rm fm}$, $\kappa=\alpha=3.316\,{\rm fm}^{-1}$, and
$A=-0.84$\,.  The circles represent the points calculated in Ref.
\protect\cite{Tiator} for the $^3$He target within the optical model.
Each curve is normalized to its own $\sigma_0$, the value of $\sigma$
at $E_\gamma=652$\,MeV.
}
\label{fig1.fig}
\end{center}
\end{figure}

\begin{figure}
\begin{center}
\unitlength=1mm
\begin{picture}(130,100)
\put(0,0){\epsfig{file=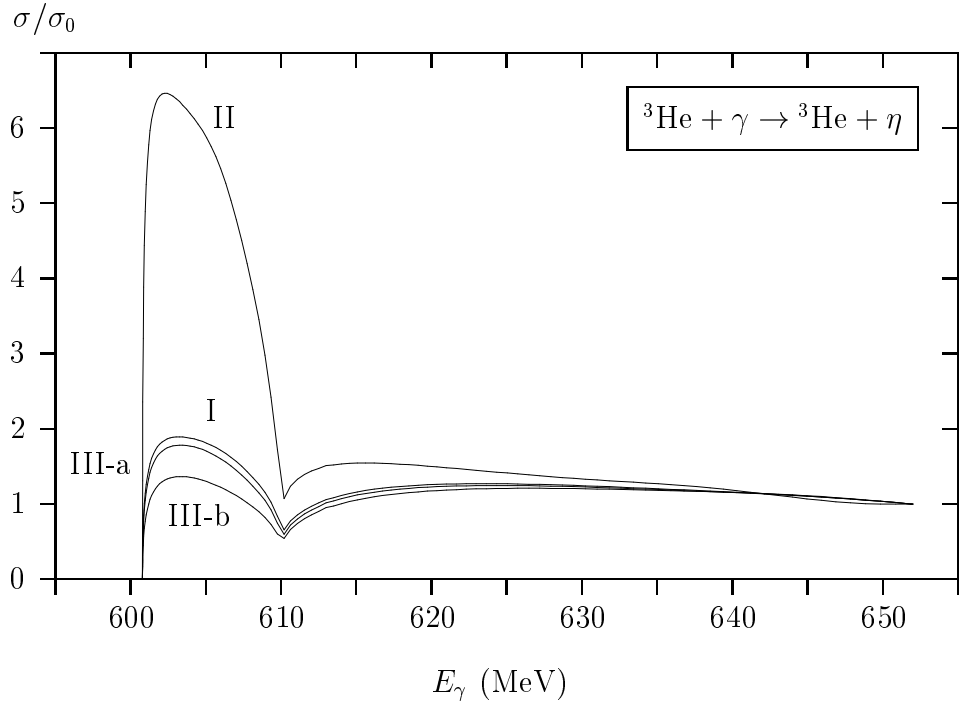}}
\end{picture}
\medskip
\caption{
Normalized cross section of the coherent $\eta$-photoproduction on $^3$He,
calculated with the four versions of $t^{\eta\eta}$
which are denoted as (I), (II), III-a, and III-b respectively.
All four curves correspond
to $\kappa=\alpha=3.316\,{\rm fm}^{-1}$  and $A=-0.84$\,.
For the curves (I) and (II) the $a_{\eta N}$ is given by
Eq.~(\ref{etaNlength}) while for the (III-a) and (III-b) curves by
Eqs.~(\ref{etaNlength1a}) and (\ref{etaNlength1b}).
}
\label{fig2.fig}
\end{center}
\end{figure}

%%%%%%%%%%%%%%%%%%%%%%%%%%%%%%%%%%%%%%%%%%%%%%%%%%%%%%%%%%%%%%%%%%%%%
Figures \ref{fig1.fig} and \ref{fig2.fig} show the results of our 
calculations for the total cross
section of the coherent process (\ref{gAAeta}). The calculations were
done for two nuclear targets, $^3$H and $^3$He, using the three
versions of $t^{\eta\eta}$ described in the section \ref{twobodyint}.
The curves corresponding to these three $T$-matrices are denoted
by (I), (II), and (III-a, III-b), respectively.

We found that the coherent $\eta$-photoproduction on these
targets is strongly enhanced in the near-threshold region as compared
to higher photon energies ($E_\gamma > 610$\,MeV).  This can be
attributed to strong final state interaction caused, for example, by a
pole of the scattering $S$-matrix, situated in the complex-energy
plane not far from the threshold energy, or in other words, to
formation of $\eta$-nucleus resonance.  In order to emphasize this finding
and to remove the insignificant but distracting differences
among different curves, we present the results in a normalized
form. Each curve shows the ratio $\sigma(E_\gamma)/\sigma_0$ with
$\sigma_0$ being the corresponding cross section at
$E_\gamma=652$\,MeV, i.e. at the energy where the near-threshold
enhancement dies out. At this energy, all the curves become flat and
are not far from each other as well as from the value of 59.812\,nb
obtained in Ref. \cite{Tiator}.  
The normalization values $\sigma_0$ are given in Table \ref{tab1.tab}.

As can be seen in Fig. \ref{fig1.fig}, the two versions of
$t^{\eta\eta}$, (I) and (II), give significantly different results
despite the fact that both of them reproduce the same $a_{\eta N}$ and
the $S_{11}$-resonance.  This indicates that the scattering of the
$\eta$ meson on the nucleons (final state interaction) is very
important in the description of the near-threshold photoproduction
process.  This conclusion is further substantiated when comparing our
curves with the corresponding points (circles) calculated for the
$^3$He target in Ref.~\cite{Tiator}. There the final state interaction
was treated using an optical potential of the first order.  It is
well-known that the first-order optical theory is not adequate at
energies near resonances. This is the reason why the calculations of
Ref.~\cite{Tiator} underestimate $\sigma$ near the threshold where,
with $a_{\eta N}=(0.75+i0.27)\,{\rm fm}$, the systems $\eta$\,${}^3$H
and $\eta$\,${}^3$He show a resonance behavior \cite{Ours23}.

Significant differences between the corresponding curves (I) and (II)
in Fig. \ref{fig1.fig} imply that two-body unitarity is important as
well.  Actually, due to the resonant character of the final state
interaction, all the details of $t^{\eta\eta}$ have strong influence
on the photoproduction cross section in the near-threshold region.
Fig. \ref{fig2.fig} where we compare the results corresponding to the
three choices of $\tau(z)$ in (\ref{tetan}), serve as another
illustration of this statement.

Since nothing is known about the parameter $\kappa$, we assume
$\kappa=\alpha$ as its basic value.  This can be motivated by the fact
that both the elastic scattering and radiative absorption
(photoproduction) of the $\eta$ meson on the nucleon go via formation
of the same $S_{11}$ resonance.  This means that at least one vertex,
namely, $\eta N\leftrightarrow S_{11}$ should be the same for both the
elastic scattering and radiative absorption.

To find out how crucial the choice of $\kappa$ is, we did two
additional calculations with $\kappa=1\,{\rm fm}^{-1}$ and
$\kappa=10\,{\rm fm}^{-1}$. We found that even with this wide
variation, the corresponding $\sigma(E_\gamma)$ curves show
practically identical enhancement of the crosss section (less than 1\%
difference). The cross section only slightly increases when the range
of the interaction becomes smaller (when $\kappa$ grows).  Therefore,
the dependence on $\kappa$ is not very strong and the choice
$\kappa=\alpha$ gives a reasonable estimate for the photoproduction
cross section.

As far as the dependence of $\sigma$ on the choice of the parameter $A
= t_{\rm n}^{\gamma \eta}/t_{\rm p}^{\gamma \eta}$ is concerned, we
found (see Sec.\ref{spinisospina}) that for $\eta$ photoproduction on
$^3$H the cross section in our model does not depend on $A$, while for
the $^3$He target the $A$-dependence is quadratic.  This means that
among these two nuclei, the helium is preferable candidate for
experimental determination of the ratio $A$.  The sign or any phase
factor of $A$, however, has no influence on the cross section if the
electromagnetic vertex is taken into account only in the first order
as it was done in our calculation.  

The cusp exhibited by all the curves at the threshold of total
nuclear break-up reflects losses of the flux into the non-coherent
channel.
%%%%%%%%%%%%%%%%%%%%%%%%%%%%%%%%%%%%%%%%%%%%%%%%%%%%%%
\acknowledgements{ The authors gratefully acknowledge financial
support from the University of South Africa, 
National Research Foundation of South Africa, the Division for
Scientific Affair of NATO (grant CRG LG 970110), and the
DFG-RFBR (grant 436 RUS 113/425/1).  One of the authors (V.B.B.)
wants to thank the Physikalisches Institut of Bonn University
for its hospitality. }

%%%%%%%%%%%%%%%%%%%%%%%%%%%%%%%%%%%%%%%%%%%%%%%%%%%%%%

%%%%%%%%%%%%%%%%%%%%%%%%%%%%%%%%%%%%%%%%%%%%%%%%%%%%%%%%
%%%%%%%%%%%%%%%    TABLES  %%%%%%%%%%%%%%%%%%%%%%%%%%%%%
%%%%%%%%%%%%%%%%%%%%%%%%%%%%%%%%%%%%%%%%%%%%%%%%%%%%%%%%
\begin{table}
\begin{tabular}{|c|c|c|c|c|c|}
\hline
target & $\sigma_0$(I), nb & $\sigma_0$(II), nb & $\sigma_0$(III-a), nb 
& $\sigma_0$(III-b), nb & $\sigma_0$(Ref. \cite{Tiator}), nb\\
\hline
$^3$H  & 49.33 & 30.58 & & & \\
\hline
$^3$He & 34.54 & 21.70 & 33.48 & 30.48 &  59.81 \\
\hline
\end{tabular}
\caption{
Values of  the total cross section of the coherent process (\ref{gAAeta})
at $E_\gamma=652$\,MeV calculated with the four versions of $t^{\eta\eta}$
which are denoted as I, II, III-a, and III-b. 
These values are used to normalize the curves shown in figures 
\ref{fig1.fig} and \ref{fig2.fig}.
}
\label{tab1.tab}
\end{table}
%%%%%%%%%%%%%%%%%%%%%%%%%%%%%%%%%%%%%%%%%%%%%%%%%%%%%%%%
%%%%%%%%%%%%%%%    FIGURES %%%%%%%%%%%%%%%%%%%%%%%%%%%%%
%%%%%%%%%%%%%%%%%%%%%%%%%%%%%%%%%%%%%%%%%%%%%%%%%%%%%%%%
%%%%%%%%%%%%%%%%%%%%%%%%%%%%%%%%%%%%%%%%%%%%%%%%%%%%%%%%%%%
\newpage

\end{document}